\documentclass[conference,letterpaper]{IEEEtran}
\usepackage[ruled,vlined,linesnumbered]{algorithm2e}
\usepackage[noend]{algpseudocode}

\usepackage[numbers,sort&compress]{natbib}

\makeatletter

\usepackage{etex}

\usepackage{zref-savepos}

\newcounter{mnote}%[page]

\def\xmarginnote{%
  \xymarginnote{\hskip -\marginparsep \hskip -\marginparwidth}}

\def\ymarginnote{%
  \xymarginnote{\hskip\columnwidth \hskip\marginparsep}}

\long\def\xymarginnote#1#2{%
\vadjust{#1%
\smash{\hbox{{%
        \hsize\marginparwidth
        \@parboxrestore
        \@marginparreset
\footnotesize #2}}}}}

\def\mnoteson{%
\gdef\mnote##1{\refstepcounter{mnote}\label{##1}%
  \zsavepos{##1}%
  \ifnum20432158>\number\zposx{##1}%
  \xmarginnote{{\color{blue}\bf $\langle$\arabic{mnote}$\rangle$}}% 
  \else
  \ymarginnote{{\color{blue}\bf $\langle$\arabic{mnote}$\rangle$}}%
  \fi%
}
  }
\gdef\mnotesoff{\gdef\mnote##1{}}
\mnoteson
\mnotesoff

\newcommand{\figref}[1]{Fig.~\ref{#1}}

\usepackage{framed}
\usepackage{subcaption}
\usepackage{comment}
\usepackage[svgnames,dvipsnames]{xcolor}
\usepackage[all]{xy}
\usepackage{tikz}
\usetikzlibrary{positioning,matrix,through,calc,arrows,fit,shapes,decorations.pathreplacing,decorations.markings,}

\tikzstyle{block} = [draw,fill=blue!20,minimum size=2em]

\usepackage{qsymbols,amssymb,mathrsfs}
\usepackage{amsmath}
\usepackage[standard,thmmarks]{ntheorem}
\theoremstyle{plain}
\theoremsymbol{\ensuremath{_\vartriangleleft}}
\theorembodyfont{\itshape}
\theoremheaderfont{\normalfont\bfseries}
\theoremseparator{}

\theoremstyle{nonumberplain}
\theoremheaderfont{\scshape}
\theorembodyfont{\normalfont}
\theoremsymbol{\ensuremath{_\blacktriangleleft}}

\theoremnumbering{arabic}
\theoremstyle{plain}
\usepackage{latexsym}
\theoremsymbol{\ensuremath{_\Box}}
\theorembodyfont{\itshape}
\theoremheaderfont{\normalfont\bfseries}
\theoremseparator{}

\theorembodyfont{\upshape}
\theoremprework{\bigskip\hrule}
\theorempostwork{\hrule\bigskip}
\usepackage[overload]{empheq} % no \intertext and \displaybreak

\let\iftwocolumn\if@twocolumn
\g@addto@macro\@twocolumntrue{\let\iftwocolumn\if@twocolumn}
\g@addto@macro\@twocolumnfalse{\let\iftwocolumn\if@twocolumn}

\mathtoolsset{showonlyrefs=false,showmanualtags}
\renewcommand{\eqref}[1]{\textup{(\refeq{#1})}} % eqref was not allowed in
\newtagform{brackets}[]{(}{)}   % new tags for equations
\usetagform{brackets}
\usepackage[Smaller]{cancel}

\PassOptionsToPackage{breaklinks,letterpaper,hyperindex=true,backref=false,bookmarksnumbered,bookmarksopen,linktocpage,colorlinks,linkcolor=BrickRed,citecolor=OliveGreen,urlcolor=Blue,pdfstartview=FitH}{hyperref}
\usepackage{hyperref}

\usepackage{graphicx,psfrag}
\graphicspath{{figure/}{image/}} % Search path of figures

\usepackage{diagbox} % \backslashbox{}{} for slashed entries
\usepackage{algorithm2e}
\usepackage{listings} 
\lstdefinelanguage{Maple}{
  morekeywords={proc,module,end, for,from,to,by,while,in,do,od
    ,if,elif,else,then,fi ,use,try,catch,finally}, sensitive,
  morecomment=[l]\#,
  morestring=[b]",morestring=[b]`}[keywords,comments,strings]
\DeclareMathAlphabet{\mathpzc}{OT1}{pzc}{m}{it}
\usepackage{upgreek} % \upalpha,\upbeta, ...
\usepackage{dsfont}  % \mathds

\def\multi@nostar#1#2{%
  \expandafter\def\csname multi#1\endcsname##1{%
    \if ##1.\let\next=\relax \else
    \def\next{\csname multi#1\endcsname}     
    \expandafter\newcommand\csname #1##1\endcsname{#2}
    \fi\next}}

\def\multi@star#1#2{%
  \expandafter\def\csname #1\endcsname##1{#2}
  \multi@nostar{#1}{#2}
}

\newcommand{\multi}{%
  \@ifstar \multi@star \multi@nostar}

\multi*{rm}{\mathrm{#1}}
\multi*{mc}{\mathcal{#1}}
\multi*{op}{\mathop {\operator@font #1}}
\multi*{ds}{\mathds{#1}}
\multi*{set}{\mathcal{#1}}
\multi*{rsfs}{\mathscr{#1}}
\multi*{pz}{\mathpzc{#1}}
\multi*{M}{\boldsymbol{#1}}
\multi*{R}{\mathsf{#1}}
\multi*{RM}{\M{\R{#1}}}
\multi*{bb}{\mathbb{#1}}
\multi*{td}{\tilde{#1}}
\multi*{tR}{\tilde{\mathsf{#1}}}
\multi*{trM}{\tilde{\M{\R{#1}}}}
\multi*{tset}{\tilde{\mathcal{#1}}}
\multi*{tM}{\tilde{\M{#1}}}
\multi*{baM}{\bar{\M{#1}}}
\multi*{ol}{\overline{#1}}

\multirm  ABCDEFGHIJKLMNOPQRSTUVWXYZabcdefghijklmnopqrstuvwxyz.
\multiol  ABCDEFGHIJKLMNOPQRSTUVWXYZabcdefghijklmnopqrstuvwxyz.
\multitR   ABCDEFGHIJKLMNOPQRSTUVWXYZabcdefghijklmnopqrstuvwxyz.
\multitd   ABCDEFGHIJKLMNOPQRSTUVWXYZabcdefghijklmnopqrstuvwxyz.
\multitset ABCDEFGHIJKLMNOPQRSTUVWXYZabcdefghijklmnopqrstuvwxyz.
\multitM   ABCDEFGHIJKLMNOPQRSTUVWXYZabcdefghijklmnopqrstuvwxyz.
\multibaM   ABCDEFGHIJKLMNOPQRSTUVWXYZabcdefghijklmnopqrstuvwxyz.
\multitrM   ABCDEFGHIJKLMNOPQRSTUVWXYZabcdefghijklmnopqrstuvwxyz.
\multimc   ABCDEFGHIJKLMNOPQRSTUVWXYZabcdefghijklmnopqrstuvwxyz.
\multiop   ABCDEFGHIJKLMNOPQRSTUVWXYZabcdefghijklmnopqrstuvwxyz.
\multids   ABCDEFGHIJKLMNOPQRSTUVWXYZabcdefghijklmnopqrstuvwxyz.
\multiset  ABCDEFGHIJKLMNOPQRSTUVWXYZabcdefghijklmnopqrstuvwxyz.
\multirsfs ABCDEFGHIJKLMNOPQRSTUVWXYZabcdefghijklmnopqrstuvwxyz.
\multipz   ABCDEFGHIJKLMNOPQRSTUVWXYZabcdefghijklmnopqrstuvwxyz.
\multiM    ABCDEFGHIJKLMNOPQRSTUVWXYZabcdefghijklmnopqrstuvwxyz.
\multiR    ABCDEFGHIJKL NO QR TUVWXYZabcd fghijklmnopqrstuvwxyz.
\multibb   ABCDEFGHIJKLMNOPQRSTUVWXYZabcdefghijklmnopqrstuvwxyz.
\multiRM   ABCDEFGHIJKLMNOPQRSTUVWXYZabcdefghijklmnopqrstuvwxyz.

\newcommand{\dotleq}{\buildrel \textstyle  .\over {\smash{\lower
      .2ex\hbox{\ensuremath\leqslant}}\vphantom{=}}}
\newcommand{\dotgeq}{\buildrel \textstyle  .\over {\smash{\lower
      .2ex\hbox{\ensuremath\geqslant}}\vphantom{=}}}

\DeclareMathOperator*{\argmin}{arg\,min}
\DeclareMathOperator*{\argmax}{arg\,max}

\newcommand{\bM}{\begin{bmatrix}}
\newcommand{\eM}{\end{bmatrix}}
\newcommand{\bSM}{\left[\begin{smallmatrix}}
\newcommand{\eSM}{\end{smallmatrix}\right]}
\renewcommand*\env@matrix[1][*\c@MaxMatrixCols c]{%
  \hskip -\arraycolsep
  \let\@ifnextchar\new@ifnextchar
  \array{#1}}

\newqsymbol{`N}{\mathbb{N}}
\newqsymbol{`R}{\mathbb{R}}
\newqsymbol{`P}{\mathbb{P}}
\newqsymbol{`Z}{\mathbb{Z}}

\newqsymbol{`|}{\mid}
\newqsymbol{`8}{\infty}
\newqsymbol{`1}{\left}
\newqsymbol{`2}{\right}
\newqsymbol{`6}{\partial}
\newqsymbol{`0}{\emptyset}
\newqsymbol{`-}{\leftrightarrow}
\newqsymbol{`<}{\langle}
\newqsymbol{`>}{\rangle}

\DeclarePairedDelimiter\abs{\lvert}{\rvert} 
\DeclarePairedDelimiter\norm{\lVert}{\rVert}

\DeclarePairedDelimiter\Set{\{}{\}}
\newcommand{\imod}[1]{\allowbreak\mkern10mu({\operator@font mod}\,\,#1)}

\newcommand{\threecols}[3]{
\hbox to \textwidth{%
      \normalfont\rlap{\parbox[b]{\textwidth}{\raggedright#1\strut}}%
        \hss\parbox[b]{\textwidth}{\centering#2\strut}\hss
        \llap{\parbox[b]{\textwidth}{\raggedleft#3\strut}}%
    }% hbox 
}

\newcommand{\reason}[2][\relax]{
  \ifthenelse{\equal{#1}{\relax}}{
    \left(\text{#2}\right)
  }{
    \left(\parbox{#1}{\raggedright #2}\right)
  }
}

\newcommand{\utag}[2]{\mathop{#2}\limits^{\text{(#1)}}}
\newcommand{\uref}[1]{(#1)}

\let\SavedDoubleVert\relax
\let\protect\relax
{\catcode`\|=\active
  \xdef\extendvert{\protect\expandafter\noexpand\csname extendvert \endcsname}
  \expandafter\gdef\csname extendvert \endcsname#1{\mskip-5mu \left.%
      \ifx\SavedDoubleVert\relax \let\SavedDoubleVert\|\fi
     \:{\let\|\SetDoubleVert
       \mathcode`\|32768\let|\SetVert
     #1}\:\right.\mskip-5mu}
}
\def\SetVert{\@ifnextchar|{\|\@gobble}% turn || into \|
    {\egroup\;\mid@vertical\;\bgroup}}
\def\SetDoubleVert{\egroup\;\mid@dblvertical\;\bgroup}

\begingroup
 \edef\@tempa{\meaning\middle}
 \edef\@tempb{\string\middle}
\expandafter \endgroup \ifx\@tempa\@tempb
 \def\mid@vertical{\middle|}
 \def\mid@dblvertical{\middle\SavedDoubleVert}
\else
 \def\mid@vertical{\mskip1mu\vrule\mskip1mu}
 \def\mid@dblvertical{\mskip1mu\vrule\mskip2.5mu\vrule\mskip1mu}
\fi

\makeatother

\usepackage{ctable}
\usepackage{fouridx}
\usepackage{framed}
\usetikzlibrary{positioning,matrix}

\usepackage{paralist}
\usepackage{enumerate}

\usepackage[normalem]{ulem}

\numberwithin{equation}{section}
\makeatletter
\@addtoreset{equation}{section}
\renewcommand{\theequation}{\arabic{section}.\arabic{equation}}
\@addtoreset{Theorem}{section}
\renewcommand{\theTheorem}{\arabic{section}.\arabic{Theorem}}
\@addtoreset{Lemma}{section}
\renewcommand{\theLemma}{\arabic{section}.\arabic{Lemma}}
\@addtoreset{Corollary}{section}
\renewcommand{\theCorollary}{\arabic{section}.\arabic{Corollary}}
\@addtoreset{Example}{section}
\renewcommand{\theExample}{\arabic{section}.\arabic{Example}}
\@addtoreset{Remark}{section}
\renewcommand{\theRemark}{\arabic{section}.\arabic{Remark}}
\@addtoreset{Proposition}{section}
\renewcommand{\theProposition}{\arabic{section}.\arabic{Proposition}}
\@addtoreset{Definition}{section}
\renewcommand{\theDefinition}{\arabic{section}.\arabic{Definition}}
\@addtoreset{Claim}{section}

\@addtoreset{Subclaim}{Theorem}
\renewcommand{\theSubclaim}{\theTheorem\Alph{Subclaim}}
\makeatother

{\endMakeFramed}

\newenvironment{ybox}{
	\setlength{\FrameSep}{1.5mm}
	\setlength{\FrameRule}{0mm}
  \MakeFramed {\FrameRestore}}%
{\endMakeFramed}

\newenvironment{gbox}{
	\setlength{\FrameSep}{1.5mm}
\setlength{\FrameRule}{0mm}
  \MakeFramed {\FrameRestore}}%
{\endMakeFramed}

{\endMakeFramed}

 {\endMakeFramed}

\usepackage{enumitem}

\usepackage{etoolbox}

\let\theparentequation\theequation
\patchcmd{\theparentequation}{equation}{parentequation}{}{}

\renewenvironment{subequations}[1][]{%              optional argument: label-name for (first) parent equation
	\refstepcounter{equation}%
	\setcounter{parentequation}{\value{equation}}%    parentequation = equation
	\setcounter{equation}{0}%                         (sub)equation  = 0
	\def\theequation{\theparentequation\alph{equation}}% 
	\let\parentlabel\label%                           Evade sanitation performed by amsmath
	\ifx\\#1\\\relax\else\label{#1}\fi%               #1 given: \label{#1}, otherwise: nothing
	\ignorespaces
}{%
	\setcounter{equation}{\value{parentequation}}%    equation = subequation
	\ignorespacesafterend
}

\newcommand*{\nextParentEquation}[1][]{%            optional argument: label-name for (first) parent equation
	\refstepcounter{parentequation}%                  parentequation++
	\setcounter{equation}{0}%                         equation = 0
	\ifx\\#1\\\relax\else\parentlabel{#1}\fi%         #1 given: \label{#1}, otherwise: nothing
}

\usepackage{blkarray}
\newcommand\circled[1]{%
  \tikz[baseline=(X.base)] 
    \node (X) [draw, dashed, shape=circle, inner sep=0] {\strut $#1$};}
\title{Agglomerative Info-Clustering}
\author{Chung Chan, Ali Al-Bashabsheh and Qiaoqiao Zhou
	\thanks{C.\ Chan (email: cchan@inc.cuhk.edu.hk),
		A.\ Al-Bashabsheh, and Q.\ Zhou are with the Institute of Network Coding at the
		Chinese University of Hong Kong, the Shenzhen Key Laboratory of
		Network Coding Key Technology and Application, China, and the
		Shenzhen Research Institute of the Chinese University of Hong
		Kong.
	}
	\thanks{The work described in this paper was supported by a grant from University Grants Committee of the Hong Kong Special Administrative Region, China (Project No. AoE/E-02/08), and supported partially by a grant from Shenzhen Science and Technology Innovation Committee (JSGG20160301170514984), the Chinese University of Hong Kong (Shenzhen), China.}
	\thanks{The work of C.\ Chan was supported in part by The Vice-Chancellor's One-off Discretionary Fund of The Chinese University of Hong Kong (Project Nos. VCF2014030 and VCF2015007), and a grant from the University
		Grants Committee of the Hong Kong Special Administrative Region,
		China (Project No. 14200714).}}

\begin{document}

\IEEEoverridecommandlockouts
%\nocite{add}
\maketitle

\begin{abstract}
An agglomerative clustering of random variables is proposed, where clusters of random variables sharing the maximum amount of multivariate mutual information are merged successively to form larger clusters. Compared to the previous info-clustering algorithms, the agglomerative approach allows the computation to stop earlier when clusters of desired size and accuracy are obtained. An efficient algorithm is also derived based on the submodularity of entropy and the duality between the principal sequence of partitions and the principal sequence for submodular functions.
\end{abstract} 

%\begin{keywords}
%multivariate mutual information; clustering; principal sequence of partitions; principal sequence; minimum norm base
%\end{keywords}

%Suggested notations: $`g_\ell$, $\mcP_{\ell}$, $I^*(\RZ_V)$, $\pzC^*(\RZ_V)$, $\pzC_{`g}(\RZ_V)$, $\pzP^*(\RZ_V)$, $`l^*_i(f)$, $\norm{x_V}$.

\section{Introduction}
\label{sec:introduction}
We consider the info-clustering paradigm proposed in \cite{chan16cluster}. It is a hierarchical
clustering of a finite set of random variables (RVs) based on the multivariate mutual information (MMI)
defined in \cite{chan15mi}. The MMI is a natural extension of Shannon's mutual information to the
multivariate case involving possibly more than two RVs. It was first proposed in
\cite{chan2008tightness} as a measure of mutual information after identifying the divergence upper
bound of the secret key agreement problem~\cite{csiszar04} to be loose in the case with helper but
tight in the no-helper case, which established an operational meaning of the MMI as the secrecy
capacity~\cite{chan10md}. The MMI was also shown to be equal to the undirected network coding
throughput~\cite{chan11isit} under the matroidal undirected network link model~\cite{chan12ud}. 

The info-clustering solution was shown in~\cite{chan16cluster} to coincide with an elegant mathematical structure called the principle sequence of
partitions (PSP)~\cite{narayanan90} of a submodular function, namely, that of the entropy function~\cite{yeung08} of the RVs to be clustered.
This leads to an algorithm~\cite[Algorithm~3]{chan16cluster} similar to \cite{nagano10} that computes the clustering solution
in $O(m^2 \op{SFM}(m))$ time, where $\op {SFM}(m)$ is the time required to minimize a submodular
function on a ground set of size $m$. In practice, however; \begin{inparaenum} 
\item one may want to obtain clusters of a desired size rather than the entire hierarchy of clusters of different sizes; and 
\item the entropy function needs to be estimated from data, which can be difficult for a large set of random
variables~\cite{wu16}.
\end{inparaenum}
These practical considerations motivate the search for an iterative info-clustering algorithm.
A divisive clustering approach was proposed in~\cite[Algorithms~1 and 2]{chan16cluster} that 
breaks down the computation by splitting the entire set of RVs successively into increasingly smaller
clusters.
However, doing so appears to be inefficient, requiring  $\Omega(m^3\op{SFM}(m))$ time in the worst case.
Furthermore, it computes the larger clusters first, the entropy function of which is more difficult
to estimate from data, and so the error may be carried forward to subsequent computations of smaller
clusters.

In this work, we propose an agglomerative info-clustering approach that aims to resolve the above
issues. The idea is to start with smaller clusters first and merge them to form larger clusters
successively. It turns out that the algorithm can be implemented more efficiently than the divisive
approach, by relating the PSP to another structure called the principal sequence
(PS)~\cite{fujishige80,fujishige05,fujishige-pp-revisited}.
A similar duality between the PSP and PS was also
used in \cite{chan16allerton} to relate info-clustering to the problem of feature selection.
The contribution of this work is the derivation of a rigorous information-theoretic interpretation
useful for the clustering problem, based on which further heuristics for estimation, approximation,
or model reduction as in~\cite{chan16cluster} can be developed.

\section{Motivation}
\label{sec:motivation}

Before a rigorous and general treatment, we introduce the problem and results informally using the following example from \cite[Figure~1a]{chan16cluster}:
Consider the following RVs defined using the uniform and independent bits $\RX_a, \RX_b, \RX_c$ and
$\RX_d$
%\begin{align}
%	&\RZ_1:=(\RX_a,\RX_d),
%	&&\RZ_2:=(\RX_a,\RX_d),
%	&\RZ_3:=\RX_a,
%	\\
%	&\RZ_4:=\RX_b,
%	&&\RZ_5:=\RX_b,
%	&\RZ_6:=\RX_c,
%\end{align}
\begin{align}
	\label{eq:eg-motivate}
	\begin{array}{lll}
		\RZ_1:=(\RX_a,\RX_d), &\RZ_2:=(\RX_a,\RX_d), &\RZ_3:=\RX_a, 
		\\
		\RZ_4:=\RX_b, &\RZ_5:=\RX_b, &\RZ_6:=\RX_c. 
	\end{array}
\end{align}
Info-clustering~\cite{chan16cluster} is a hierarchical clustering approach based on a multivariate
measure of the information shared among (multiple) RVs.
As will be discussed more precisely in a subsequent section, info-clustering provides clusters for 
different thresholds $`g\in \mathbb{R}$, where a cluster is an inclusion-wise maximal subset of RVs
that share more than $`g$ amount of information. (We do not regard a singleton as a cluster.)
Since the correlation structure of the RVs in \eqref{eq:eg-motivate} is simple, let us for the
moment define such a measure of information among $\RZ_B$, for any $B\subseteq \{1,\ldots,6\}$ with $|B|\geq 2$, as the number of
bits shared by $\RZ_B$ and denote it by $I(\RZ_B)$.
Then we have
\begin{itemize}
	\item $I(\RZ_{\{1,\ldots,6\}}) = 0$ since, e.g., $\RZ_{\{1,\ldots,5\}}$ and $\RZ_6$ share no bits. 
		(Similarly, $I(\RZ_{B\cup \{6\}}) = 0$ for $\emptyset \neq B \subseteq \{1,\cdots,5\}$;
		$I(\RZ_{\{1,4\}})=0$; etc.)
	\item
		$I(\RZ_{\{1,2,3\}}) = 1$
		since $\RZ_1$, $\RZ_2$, and $\RZ_3$ share the bit $\RX_a$.
		(Similarly, $I(\RZ_{\{1,3\}}) = 1 = I(\RZ_{\{2,3\}})$.)
		We also have
		$I(\RZ_{\{4,5\}}) = 1$
		since $\RZ_4$ and $\RZ_5$ share the bit $\RX_b$.
	\item $I(\RZ_{\{1,2\}}) = 2$ since $\RZ_1$ and $\RZ_2$ share the two bits $\RX_a$ and
		$\RX_d$. (This is the only set of RVs sharing more than one bit.)
	\item There is no set of RVs that share more than two bits.
\end{itemize}
%, i.e., for all $B\subseteq V$ with $|B| \geq 2$,
%\begin{align}
%	I(\RZ_B):=\left\{
%		\begin{array}{ll}
%			0, & B \ni 6 \\
%			1, & B = \{1,\ldots,6\} \\
%		\end{array}
%		\right.
%\end{align}
Hence, for all $`g \in \mathbb{R}$, the collection of clusters at threshold $`g$, denoted as
$\pzC_{`g}(\RZ_{\{1,\ldots,6\}})$, is given by~\cite[Figure~1b]{chan16cluster}
\begin{align}
	\label{eq:eg-clusters}
	\pzC_{`g}(\RZ_{\{1,\ldots,6\}}) = \left\{
		\begin{array}{ll}
			\{\{1,\ldots,6\}\}, 		& `g < 0 \\
			\{\{1,2,3\},\{4,5\}\}, 	& `g \in [0,1) \\
			\{\{1,2\}\},			 	& `g \in [1,2) \\
			\emptyset,				 	& `g \geq 2.
		\end{array}
		\right.
\end{align}
%where by a cluster at threshold $`g$ we mean an inclusion-wise maximal collection of (at least two) RVs   
%sharing more than $`g$ bits. %(The precise definition will be given in \eqref{}.)
For instance, for any $`g<0$, there is only one cluster, namely, the entire set of RVs since any subset
containing at least two RVs satisfies the threshold constraint and the entire set $\{1,\ldots,6\}$
is trivially the maximal one. For $`g =0$, the threshold constraint dictates that we seek
collections of RVs that share a strictly positive number of bits, i.e., one or two bits
in this example.
Of such sets, it is easy to verify that the maximal ones are $\{1,2,3\}$ and $\{4,5\}$.
This remains to be the case for any $`g\in [0,1)$. For $`g=1$, we seek collections of RVs that share
	more than one bits, i.e., two bits in this example. The set $\{1,2\}$ is the only such set, and
	so, it is trivially maximal. This remains to be the case for $`g\in [1,2)$. Finally, for
		$`g\geq 2$, we have no clusters since no set of RVs share more than two bits of information.

		The reader may have observed the following hierarchical structure: Starting with the cluster
$\{1,\ldots,6\}$ at sufficiently small threshold $`g$, the cluster breaks into two smaller clusters
$\{1,2,3\}$ and $\{4,5\}$, where each is a maximal subset that has more shared information bits
than the original cluster. Continuing in this fashion, the cluster $\{1,2,3\}$
		breaks into the smaller cluster $\{1,2\}$, at which point no further breakage is possible and 
		all clusters have been found.
		More generally, under a proper choice of the multivariate information measure, the
		hierarchical structure of the clusters persists, thereby allowing a divisive algorithm
		for finding the clusters \cite[Algorithm~1]{chan16cluster}. The algorithm starts with the entire set of RVs then
		proceeds iteratively to break the clusters into smaller and smaller disjoint clusters.

		In this work, we propose the reverse procedure for computing the clusters. Namely, an
		\emph{agglomerative} approach where RVs gradually group into larger and larger clusters. In
our example, starting with the singletons for sufficiently large value of $`g$, merge $\{1\}$ and $\{2\}$ into a cluster since such
		a merging results in the maximal subset $\{1,2\}$ with the maximum number of shared bits. Continue to
merge $\{1,2\}$ with $\{3\}$ and merge $\{4\}$ with $\{5\}$ to form the maximal subsets $\{1,2,3\}$ and
$\{4,5\}$ with the
		second largest number of shared bits. Continue in the same way to merge $\{1,2,3\}$, $\{4,5\}$, and $\{6\}$ into the
		cluster $\{1,\ldots,6\}$, at which point no further merging is possible and all clusters have
		been found.
	
		The practicality of an agglomerative approach in comparison to a divisive one is that the
former identifies clusters of larger amount of shared bits first. A larger amount of shared bits can be estimated from data
		more accurately, and so it serves as a stronger base for the subsequent estimation for finding
		clusters of smaller amount of shared bits. Moreover, as we will subsequently show, when the information measure is chosen to be the MMI in \cite{chan15mi}, the agglomerative approach is computationally more efficient compared to the divisive one.
%\textcolor{red}{---probably talk about factor $n$ faster, hence, closing the gap between the iterative and non-iterative approaches toward clusters\ldots}

\section{Problem formulation}
\label{sec:problem}
Given a random vector $\RZ_V:=(\RZ_i \mid i \in V)$ where $V$ is
an ordered finite set of $|V|>1$ RVs. The set of clusters at any threshold $`g\in `R$ is defined in \cite{chan16cluster} as
\begin{align}
	\pzC_{`g}(\RZ_V):=\op{maximal}\{B\subseteq V \mid |B| > 1, I(\RZ_B) > `g\},\label{eq:clusters}
\end{align}
where $\op{maximal}\pzF := \{B \in \pzF \mid \nexists B' \in \pzF, B \subsetneq B' \}$
denotes the collection of inclusion-wise maximal elements of any collection $\pzF$ of subsets, and 
$I(\RZ_B)$ is a multivariate information measure satisfying
\begin{align}
	\label{eq:imunion}
	I(\RZ_{B_1 \cup B_2}) \geq \min\{I(\RZ_{B_1}), I(\RZ_{B_2})\}
\end{align}
for all $B_i \subseteq V$ with $\abs{B_i}>1$, $i\in \Set {1,2}$ and $B_1\cap B_2 \neq \emptyset$.
It was shown in \cite[Theorem~3]{chan16cluster} %\cite[Theorem~2.3]{chan16cluster}
 that the clusters form a laminar family, i.e., 
for any $`g' \leq `g''$,
$C' \in \pzC_{`g'}(\RZ_V)$,
and
$C'' \in \pzC_{`g''}(\RZ_V)$,
we have
\begin{align}
	\label{eq:laminar}
	C' \cap C'' = `0 \kern1em \text{or}\kern1em  C' \supseteq C''.
\end{align}
In particular, clusters at the same threshold must be disjoint.
Consequently, the clustering solution can be characterized as follows by set partitions of $V$, the collection of which is denoted as $\Pi(V)$.

\begin{Proposition}[\mbox{\cite[Theorems~1 and 4]{chan16cluster}}]%[\mbox{\cite[Theorems~2.1 and 2.4]{chan16cluster}}]
	\label{prop:clusters}
	The property~\eqref{eq:imunion} implies that the clustering solution~\eqref{eq:clusters} satisfies
	\begin{align*}
		\pzC_{`g}(\RZ_V) &= \mcP_{\ell}\backslash \{\{i\}\mid i\in V\} \kern1em \text {for } `g \in [`g_\ell,`g_{\ell+1}], 0\leq \ell \leq N
	\end{align*}
	where $N$ is a positive integer; $`g_0:=-`8$ and $`g_{N+1}:=`8$ for convenience;
		\begin{align}
			\label{eq:laminar-gamma-seq}
			-\infty< `g_1 < \cdots < `g_N < \infty
		\end{align}
		is a sequence of distinct critical values from $`R$ (consisting of the thresholds at which the set of clusters changes); and 
		\begin{align}
			\label{eq:laminar-partitions-seq}
			\kern-1em \mcP_0=\{V\} \succ \mcP_1 \succ \cdots \succ \mcP_{N-1} \succ
			\mcP_{N}=\{\{i\}:i\in V\}\kern-.5em
		\end{align}
		is a sequence of increasingly finer partitions of $V$ from $\Pi(V)$, where $\mcP' \succeq \mcP$ means that
		\begin{align}
			\label{eq:finer}
			\forall C \in \mcP,  \exists  C' \in \mcP' : C \subseteq C',
		\end{align}
		and ``$\succ$" denotes the strict inequality  (i.e., the inclusion above is strict for at least
		one $C \in \mcP$).
\end{Proposition}
%To emphasize their dependence on $\RZ_V$, we will subsequently denote $N$, $`g_\ell$, and $\mcP_{\ell}$ as $N(\RZ_V)$, $`g_{\ell}(\RZ_V)$, and $\mcP_{\ell}(\RZ_V)$, respectively. For convenience, $`g_{N}(\RZ_V)$ will denote $`g_{N(\RZ_V)}(\RZ_V)$ and similarly $\mcP_{N}(\RZ_V)$ will denote $\mcP_{N(\RZ_V)}(\RZ_V)$.

With the laminar structure, a divisive clustering algorithm was given in \cite{chan16cluster} to 
compute $`g_\ell$ and $\mcP_{\ell}$ by iteratively producing finer partitions from coarser ones, i.e., from
$\ell=1$ to $\ell =N$.
The work in \cite{chan15mi} considered a particularly meaningful multivariate information measure,
namely, the \emph{multivariate mutual information} (MMI) defined as
\begin{subequations}
	\label{eq:mmi}
\begin{align}
	I(\RZ_V)&:=\min_{\mcP\in \Pi'(V)} I_{\mcP}(\RZ_V), \kern1em \text{where} \label{eq:I}\\
	I_{\mcP}(\RZ_V)&:= \frac{1}{|\mcP|-1}`1[\sum\nolimits_{C\in \mcP}H(\RZ_C) - H(\RZ_V) `2]\label{eq:IP}
\end{align}
\end{subequations}
and $\Pi'(V):=\Pi(V)`/\Set{\Set {V}}$ is the set of all partitions of $V$ into at least two non-empty disjoint subsets.
The MMI was first proposed as a measure of mutual information in \cite{chan2008tightness} and was later shown in \cite{chan15mi} to satisfy
\eqref{eq:imunion}, together with various other properties that naturally extend those of Shannon's mutual
information to the multivariate case. The MMI  was also later referred to in \cite{CIT-072} as ``shared information." An important result that inspired the info-clustering paradigm~\cite{chan16cluster} is:
%Roughly speaking, the following result from \cite{chan} asserts that under the MMI measure, the clustering solution (for a particular threshold) coincides with a specific optimal partition to \eqref{eq:mmi}.
%(Stated differently, one of the defining partitions of the MMI also defines the clustering solution
%under the MMI for a particular threshold.)
\begin{Proposition}[\mbox{\cite[Theorems~5.2 and 5.3]{chan15mi}}]
	\label{prop:fund:cluster}
	The optimal partitions achieving the MMI in~\eqref{eq:mmi} together with the trivial partition $\{V\}$ form a
	lattice w.r.t.\ \eqref{eq:finer}. The minimum/finest optimal partition, denoted by
	$\pzP^{*}(\RZ_V)$, satisfies
	\begin{align*}
		\pzP^{*}(\RZ_V) \backslash \{\{i\}\mid i\in V\} = \pzC_{I(\RZ_V)}(\RZ_V),
	\end{align*}
	where $\pzC_{`g}$ is defined in \eqref{eq:clusters} using the MMI.
	In particular,  $`g_1 = I(\RZ_V)$ and $\mcP_{1} = \pzP^{*}(\RZ_V)$ in Proposition~\ref{prop:clusters}.
\end{Proposition}
Together with the laminar structure in Proposition~\ref{prop:clusters}, it can be argued that any
algorithm for computing $\pzP^{*}(\RZ_V)$ can be applied iteratively for divisive info-clustering as
in \cite[Algorithm~2]{chan16cluster}. However, doing so appears to be less efficient compared to the
alternative approach in \cite[Algorithm~3]{chan16cluster} that computes all the clusters but not
in any particular order. Nevertheless, in practice it is 
desirable to have an iterative approach that can stop when further computation is not of
interest or is meaningless due to errors in estimating or approximating the entropies from data. 

%%%%%%%%%%%%%%%%%%%%%%%%%%%%%%%%%
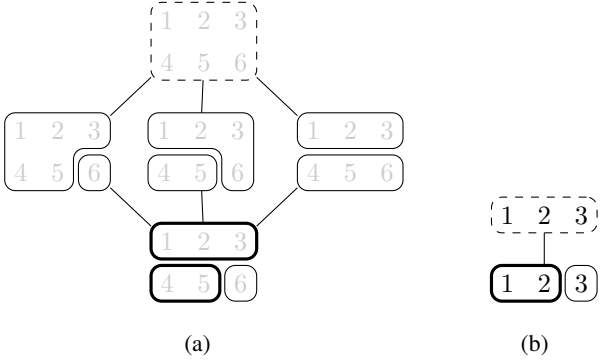
\begin{figure}
	\begin{center}
		\subcaptionbox{ \label{fig:eg-div-zv} }{
		\def\thickness{very thick}
		\begin{tikzpicture}[
		group/.style={fill opacity=.2, inner sep=0, outer sep=0, rounded corners},
		every node/.style={rounded corners, text opacity=1, fill opacity=.2}
		]

		\matrix(p0)at(2,-2)[matrix of math nodes, ampersand replacement=\&,
		row sep=1mm,
		column sep=.8mm,
		every cell/.style={anchor=base west}]{
			1 \& 2\& 3 \\
			4 \& 5\& 6 \\
			\\};
		\def\dist{1.8}
		\def\distx{.3cm}
		\def\disty{.1cm}
		\matrix(pl0)[below left = \disty and \distx of p0,
		matrix of math nodes, ampersand replacement=\&,
		row sep=1mm,
		column sep=.8mm,
		every cell/.style={anchor=base west}]{
			1 \& 2\& 3 \\
			4 \& 5\& 6 \\
			\\};
		\matrix(pm0)[below = \disty of p0,
		matrix of math nodes, ampersand replacement=\&,
		row sep=1mm,
		column sep=.8mm,
		every cell/.style={anchor=base west}]{
			1 \& 2\& 3\& \\
			4 \& 5\& 6\& \\
			\\};
		\matrix(pr0)[below right = \disty and \distx of p0,
		matrix of math nodes, ampersand replacement=\&,
		row sep=1mm,
		column sep=.8mm,
		every cell/.style={anchor=base west}]{
			1 \& 2\& 3\& \\
			4 \& 5\& 6\& \\
			\\};
		\matrix(p1)[below = \disty of pm0,
		matrix of math nodes, ampersand replacement=\&,
		row sep=1mm,
		column sep=.8mm,
		every cell/.style={anchor=base west}]{
			1 \& 2\& 3 \\
			4 \& 5\& 6 \\
			\\};
		
		% clustering of Z in PLP
		\node[dashed, draw, group, fill=none, fit=(p0-1-1)(p0-2-3)]{};
		%			%
		\node(pp0)[draw, group, fill=none, fit=(pl0-2-3)]{};
		\draw[rounded corners](pl0-1-1.north west)--(pl0-1-3.north east)--(pl0-1-3.south
		east)--(pl0-1-2.south east)--(pl0-2-2.south east)--(pl0-2-1.south west)-- cycle;
		%			%
		\node(pp0)[draw, group, fill=none, fit=(pm0-2-1)(pm0-2-2)]{};
		\draw[rounded corners](pm0-1-1.north west)--(pm0-1-3.north east)--(pm0-2-3.south east)
		--(pm0-2-3.south west)--(pm0-1-3.south west)--(pm0-1-1.south west)-- cycle;
		%			%
		\node[draw, group, fill=none, fit=(pr0-1-1)(pr0-1-3)]{};
		\node[draw, group, fill=none, fit=(pr0-2-1)(pr0-2-3)]{};
		%			%
		\draw(p0-2-1)--(pl0-1-3);
		\draw(p0-2-2)--(pm0-1-2);
		\draw(p0-2-3)--(pr0-1-1);
		%			%
		\draw(p1-1-1)--(pl0-2-3);
		\draw(p1-1-2)--(pm0-2-2);
		\draw(p1-1-3)--(pr0-2-1);
		\node(pp0)[draw, \thickness, group, fill=none, fit=(p1-1-1)(p1-1-3)]{};
		\node(pp0)[draw, \thickness, group, fill=none, fit=(p1-2-1)(p1-2-2)]{};
		\node(pp0)[draw, group, fill=none, fit=(p1-2-3)]{};
		%			%
		\end{tikzpicture}
		}%\hfill
		\subcaptionbox{\label{fig:eg-div-z123}}[3cm]{
			\def\thickness{very thick}
			\begin{tikzpicture}[
			group/.style={fill opacity=.2, inner sep=0, outer sep=0, rounded corners},
			every node/.style={rounded corners, text opacity=1}
			]

			\matrix(p0)at(2,-2)[matrix of math nodes, ampersand replacement=\&,
			row sep=1mm,
			column sep=.8mm,
			every cell/.style={anchor=base west}]{
				1 \& 2\& 3 \\
				\\};
			\def\dist{1.8}
			\def\distx{.3cm}
			\def\disty{.1cm}
			\matrix(p1)[below = \disty of p0,
			matrix of math nodes, ampersand replacement=\&,
			row sep=1mm,
			column sep=.8mm,
			every cell/.style={anchor=base west}]{
				1 \& 2\& 3 \\
				\\};
			
			% clustering of Z in PLP
			\node[dashed, draw, group, fill=none, fit=(p0-1-1)(p0-1-3)]{};
			%			%
			\node[draw, \thickness, group, fill=none, fit=(p1-1-1)(p1-1-2)]{};
			\node[draw, group, fill=none, fit=(p1-1-3)]{};
			%			%
			\draw(p0-1-2)--(p1-1-2);
			\end{tikzpicture}
		}
%		\hfill
%		\subcaptionbox{$\pzP^{*}(\RZ_{\Set{4,5}})$ \label{fig:eg-div-z45}}[3cm]{
%			\input{dtl/dtl-45.tex}
%		}
	\end{center}
	\caption{Optimal partitions of: (a) $I(\RZ_V)$ and (b) $I(\RZ_{\Set{1,2,3}})$. In each case, the
fundamental partition is the one at the bottom, where the associated clusters are circled with
thick lines.}
\label{fig:eg-div}
\end{figure}
\begin{example}
	As an illustration of Proposition~\ref{prop:fund:cluster} and the divisive approach, consider our
	motivating example.
	%is shown in \figref{fig:eg-div} using our motivating example.
	The finest optimal partition
	$\pzP^{*}(\RZ_V) = \Set{\Set{1,2,3},\Set{4,5},\Set{6}}$ is shown in \figref{fig:eg-div-zv}. For
	completion, the figure also shows the lattice of optimal partitions stated in the proposition,
	where the trivial partition $\Set{V}$ is indicated using a dashed line.
	Similarly, \figref{fig:eg-div-z123} shows the optimal partitions of
	$\RZ_{\Set{1,2,3}}$.
	Since $I(\RZ_{V}) = 0$ and $I(\RZ_{\Set{1,2,3}}) = 1$, Proposition~\ref{prop:fund:cluster}
	asserts that $\pzC_{0} = \Set{\Set{1,2,3},\Set{4,5}}$ and $\pzC_{1} = \Set{\Set{1,2}}$, in agreement
	with \eqref{eq:eg-clusters}.
	Assuming an algorithm for computing the
	finest optimal partition, the divisive algorithm starts by computing 
	$\pzP^{*}(\RZ_V) = \Set{\Set{1,2,3},\Set{4,5},\Set{6}}$ (from which $I(\RZ_V)$ is readily
	available), declares the non-singleton elements as clusters at threshold $I(\RZ_V)$,
	%(the set $\Set{1,2,3}$ in our example). 
	%declare the non-singleton elements as
	%the set of clusters at $I(\RZ_V)$,
	and proceeds iteratively by picking any cluster (of size larger than two) and computing its
	finest optimal partition, etc. In our example, there is only one cluster, the set $\Set{1,2,3}$
	at threshold $0$. The finest optimal partition of $\RZ_{\Set{1,2,3}}$ is shown in
	Fig.~\ref{fig:eg-div-z123}, which results in the cluster $\Set{1,2}$ at threshold
	$I(\RZ_{\Set{1,2,3}}) = 1$. After this, the divisive algorithm terminates in this example.
\end{example}
%%%%%%%%%%%%%%%%%%%%%%%%%%%%%%%%%

\begin{algorithm}
	\caption{Agglomerative info-clustering.}
	\label{alg:aic}
	\SetKwFunction{FirstClusters}{FirstClusters}
	\SetKwFunction{Fuse}{Fuse}
	\SetKwProg{myfn}{function}{:}{end}
	\DontPrintSemicolon
	\SetAlgoLined
	\KwData{Statistics of $\RZ_V$ sufficient for calculating the entropy function $h(B)$ for
		$B\subseteq V:=\{1,\ldots,n\}$.}
	\KwResult{The arrays \DataSty{L} and \DataSty{PSP} contain the $`g_{\ell}$'s and $\mcP_{\ell}$'s
		in Proposition~\ref{prop:clusters}. More precisely,
		for $1\leq \ell \leq N$, the entries of the arrays \DataSty{PSP} and
		\DataSty{L} are 
		\DataSty{PSP}$[s] = \mcP_{\ell}(\RZ_V)$ and
		\DataSty{L}$[s] = `g_{\ell}(\RZ_V)$
		for $s = |\mcP_{\ell}(\RZ_V)|$, 
		and are null otherwise.}
	\BlankLine
	\DataSty{L}, \DataSty{PSP}$\leftarrow$ empty arrays, each of length $n$;\;
	\DataSty{PSP}$[n]$ $\leftarrow \{\{i\}:i\in V\}$, $s\leftarrow n$;\;
	\While{$s>1$}{
		$(`g,\mcP')\leftarrow$ $\Fuse($\DataSty{PSP}$[l])$;\;
		\DataSty{L}$[s]\leftarrow `g$;\;
		$s\leftarrow |\mcP'|$;\;
		\DataSty{PSP}$[s]\leftarrow \mcP'$;\;
	}
	% 	\BlankLine
	%   	\myfn{\Fuse{$\mcP$}}{
	%			assume $\mcP=\mcP_{\ell}$ for $1\leq \ell\leq N$ in Proposition~\ref{prop:clusters};\;
	%			\KwRet $(`g,\mcP_{\ell}) \leftarrow (`g_{\ell},\mcP_{\ell-1})$;
	%   	}
	%	\textcolor{red}{---needs revision, particularly the function fuse.}
\end{algorithm}

Instead of the divisive approach, we consider here an agglomerative approach shown in Algorithm~\ref{alg:aic} that computes $`g_{\ell}$'s and $\mcP_{\ell}$'s
iteratively from $\ell=N$ down to $\ell=1$. We will give an efficient implementation of
the subroutine \Fuse in Algorithm~\ref{algo:fuse} that computes $`g_\ell$ and $\mcP_{\ell-1}$ from
$\mcP_{\ell}$ for any $\ell$. % $\ell = 1, \dots, N$.
In particular, we will show that it suffices to compute
\begin{align}
	I^{*}(\RZ_V)&:=\max\{I(\RZ_{B})\mid B\subseteq V, |B| > 1\} \kern1em \text{ and}\label{eq:I*}\\
	\begin{split}
	\pzC^{*}(\RZ_V)&:=\op{maximal}\{B\subseteq V \mid \abs{B} > 1, \\
	&\kern8em I(\RZ_B) = I^{*}(\RZ_{V})  \},
	\end{split}\label{eq:C*}
\end{align}
which are clearly the last critical value $`g_{N}$ and the non-singleton elements of the second last
partition $\mcP_{N-1}$ respectively.

\section{Preliminaries}
\label{sec:prelim}
The ability to compute info-clustering solution efficiently stems from the submodularity of entropy~\cite{fujishige78}, or equivalently the fact that mutual information is non-negative~\cite{shannon48}. More precisely, by denoting the entropy function of $\RZ_V$ as 
\begin{align}
	h(B):=H(\RZ_B) \kern1em \text{for } B\subseteq V, \label{eq:h}
\end{align}
(where the dependency on $\RZ_V$ is implicit for convenience,) \emph{submodularity} of $h$ means that
\begin{align}
	h(B_1)+h(B_2)\geq h(B_1\cup B_2)+h(B_1\cap B_2)  \label{eq:submodular}
\end{align}
for all $B_1,B_2\subseteq V$.
$h$ is also said to be \emph{normalized} as $h(`0)=0$, and non-decreasing as $h(B')\leq h(B)$ whenever $B' \subseteq B \subseteq V$. %(All these properties are referred to as Shannon-type inequalities~\cite{yeung08})
In combinatorial optimization~\cite{schrijver02}, submodularity is well-known to give rise to polynomial-time solutions. 
The info-clustering problem, in particular, relies on the following closely related polynomial-time solvable structures.

\subsection{Principal sequence of partitions}
\label{sec:PSP}

For any real number $`g\in `R$, define the residual entropy function~\cite{chan15mi} of a random vector $\RZ_V$ as
\begin{align}
	h_{`g}(B)&:= h(B)-`g \kern2.5em \text{for $B\subseteq V$}. \label{eq:residualH}
\end{align}
The residual entropy function is also submodular and its Dilworth truncation evaluated at $V$ is defined as~\cite{schrijver02}
\begin{subequations}
	\label{eq:DT}
\begin{align}
		\hat{h}_{`g}(V)&:= \min_{\mcP\in \Pi(V)} h_{`g}[\mcP] \kern1em\text{where}\label{eq:DT:1}\\
		h_{`g}[\mcP] &:= \sum\nolimits_{C\in \mcP} h_{`g}(C). \label{eq:DT:2} 
\end{align} 
\end{subequations}
\begin{Proposition}[\mbox{\cite{narayanan90}}]
	\label{prop:plp}
	Submodularity~\eqref{eq:submodular} of $h$~\eqref{eq:h} implies that the set of optimal
	partitions to the Dilworth truncation~\eqref{eq:DT:1} forms a lattice, called the Dilworth
	truncation lattice, with respect to the partial order~\eqref{eq:finer}. Furthermore, if $\mcP'$
	and $\mcP''$ are the optimal partitions for $`g'$ and $`g''$ respectively, then $`g'<`g''$
	implies $\mcP'\succeq \mcP''$.
\end{Proposition}
In particular, the minimum/finest optimal partition exists and characterizes the info-clustering solution as follows:
% the corresponding threshold. %and $\Pi(V)$ is the collection of all partitions of $V$ into non-empty subsets.
\begin{Proposition}[\mbox{\cite[Corollary~2]{chan16cluster}}]%[\mbox{\cite[Corollary~3.1]{chan16cluster}}]
	\label{prop:psp}
	For a finite set $V$ with size $\abs{V}>1$ and a random vector $\RZ_V$,
	\begin{align}
		\begin{split}
	\pzC_{`g}(\RZ_V)
	&=`1[\min \Set{\mcP\in \Pi(V) \mid h_{`g}[\mcP]=\hat{h}_{`g}(V)}`2]\\ &\kern4em \big\backslash`1\{\Set{i}\mid i\in V`2\}, 
	   \end{split}
	\end{align}
namely, the non-singleton elements of the finest optimal partition to the Dilworth truncation~\eqref{eq:DT:1}.
	%where $\hat{h}_{`g}$ is the Dilworth truncation~\eqref{eq:DT:2} of the residual entropy function~\eqref{eq:residualH} of $\RZ_V$. 
\end{Proposition}

Hence, in Proposition~\ref{prop:clusters}, the sequence of $\mcP_{\ell}$ for $\ell$ from $1$ to $N$ with the corresponding $`g_\ell$ also characterizes the minimum optimal partitions to~\eqref{eq:DT:1} for all $`g\in `R$, and is known as the principal sequence of partitions (PSP) of $h$, introduced in~\cite{narayanan90}. 

%%%%%%%%%%%%%%%%%%%%%%%%%%%%
\begin{figure}
	\begin{center}
%		\subcaptionbox{PLP \label{fig:eg-plp}}{
%			\input{dtl/plp.tex}
%		}%\hfill
		\subcaptionbox{$\hat{h}_{`g}(V)$ \label{fig:eg-dt-fn}}{
			\hspace{-.8cm}
			{\def\u{1.2}
				\def\sx{0.9}
				\def\left{left}
				\def\right{right}
				\def\plabel{-4}
				\tikzstyle{point}=[draw,solid,red,thick,circle,minimum size=.2em,inner sep=.0em, outer sep=.2em]
				\begin{tikzpicture}[remember picture,x=1.4em,y=1.4em,>=latex, every node/.style={font=\scriptsize}]
				\draw[->] (0,-6.5*\u) -- (0,6.5*\u) node (y) [label=right:$\hat{h}_{`g}(V)$] {};
				\draw[->] (-1*\sx*\u,0) -- (3*\sx*\u,0) node [label=below:$`g$] {};
				\foreach \i/\ya/\xa/\yb/\xb/\lp/\ld/\lb in {
					1/5.0/-1.0*\sx/4/0*\sx/above \left/.7em/{$\kern0em h_{`g}[\Set{\Set{1,\dots,6}}]=4-`g$}, 
					2/4/0*\sx/1/1*\sx/\left/0em/{$h_{`g}[\Set{\Set{1,2,3},\Set{4,5},\Set{6}}]=4-3`g$},
					3/1/1*\sx/-4/2*\sx/\left/0em/{$h_{`g}[\Set{\Set{1,2},\Set{3},\Set{4},\Set{5},\Set{6}}]=6-5`g$},
					4/-4/2*\sx/-6.5/2.35*\sx/below \left/0em/{$h_{`g}[\Set{\Set{1},\Set{2},\Set{3},\Set{4},\Set{5},\Set{6}}]=8-6`g$}    
					%4/-4/2/-7/2.5/left/{$h_{`g}[\Set{\Set{1},\Set{2},\Set{3},\Set{4},\Set{5},\Set{6}}]=8-6`g$}    
				}
				\draw[dashed] (\xa*\u,\ya*\u)  to node [inner sep=0,outer sep=0,label={[label distance=\ld]\lp:{\scriptsize\lb}}] {}  (\xb*\u,\yb*\u);
				%\path (0*\sx*\u,4*\u) node (1) [point,red,thick,label=\right:{\scriptsize$p_1$}] {};
				%\path (1*\sx*\u,1*\u) node (2) [point,red,thick,label={[label distance=0em]\right:{\scriptsize$p_2$}}] {};
				%\path (2*\sx*\u,-4*\u) node (3) [point,red,thick,label=\right:{\scriptsize$p_3$}] {};
				\draw[dashed,->](\plabel*\sx*\u,4*\u) node[below]{$0=I(\RZ_{\Set{1,\dots,6}})$} -- (0*\sx*\u,4*\u)node(1)[point]{} -- (0*\sx*\u,0);
				\draw[dashed,->](\plabel*\sx*\u,1*\u) node[above]{$1=I(\RZ_{\Set{1,2,3}})$}node[below]{$=I(\RZ_{\Set{4.5}})$} -- (1*\sx*\u,1*\u)node(2)[point]{} -- (1*\sx*\u,0*\u);
				\draw[dashed,->](\plabel*\sx*\u,-4*\u) node[above]{$2=I(\RZ_{\Set{1,2}})$} -- (2*\sx*\u,-4*\u) node(3)[point]{}-- (2*\sx*\u,0);
				\draw[-,thick,blue] (-1*\sx*\u,5*\u)--(1)--(2)--(3)--(2.35*\sx*\u,-6.5*\u);
				\end{tikzpicture}}
		}%\hfill
		\subcaptionbox{PSP \label{fig:eg-psp}}{
			\def\thickness{very thick}
			\begin{tikzpicture}[remember picture,rotate=0,
			group/.style={fill opacity=.2, inner sep=0, outer sep=0, rounded corners},
			every node/.style={rounded corners, text opacity=1, transform shape}
			]
			\def\position{below}
			\matrix(p0)at(0,0)[matrix of math nodes, ampersand replacement=\&,
			row sep=1mm,
			column sep=.8mm,
			every cell/.style={anchor=base west}]{
				1 \& 2\& 3 \\
				4 \& 5\& 6 \\
				\\};
			\def\dist{1.8}
			\def\distx{.3cm}
			\def\disty{.8cm}
			\matrix(p1) [\position = 0.9*\disty of p0,
			matrix of math nodes, ampersand replacement=\&,
			row sep=1mm,
			column sep=.8mm,
			every cell/.style={anchor=base west}]{
				1 \& 2\& 3 \\
				4 \& 5\& 6 \\
				\\};
			\matrix(p2)[\position = 1.1*\disty of p1,
			matrix of math nodes, ampersand replacement=\&,
			row sep=1mm,
			column sep=.8mm,
			every cell/.style={anchor=base west}]{
				1 \& 2\& 3 \\
				4 \& 5\& 6 \\
				\\};
			\matrix(p3)[\position = 1.1*\disty of p2,
			matrix of math nodes, ampersand replacement=\&,
			row sep=1mm,
			column sep=.8mm,
			every cell/.style={anchor=base west}]{
				1 \& 2\& 3 \\
				4 \& 5\& 6 \\
				\\};
			% clustering of Z in PLP
			\node(p0)[draw, group, fill=none, fit=(p0-1-1)(p0-2-3)]{};
			\node[draw, \thickness, group, fill=none, fit=(p1-1-1)(p1-1-3)]{};
			\node[draw, \thickness, group, fill=none, fit=(p1-2-1)(p1-2-2)]{};
			\node[draw, group, fill=none, fit=(p1-2-3)(p1-2-3)]{};
			\node(p1)[group, fill=none, fit=(p1-1-1)(p1-2-3)]{};
			\node[draw, group, \thickness, fill=none, fit=(p2-1-1)(p2-1-2)]{}; \node[draw, group, fill=none, fit=(p2-1-3)]{};
			\node[draw, group, fill=none, fit=(p2-2-1)]{};
			\node[draw, group, fill=none, fit=(p2-2-2)]{};
			\node[draw, group, fill=none, fit=(p2-2-3)]{};
			\node(p2)[group, fill=none, fit=(p2-1-1)(p2-2-3)]{};
			\node[draw, group, fill=none, fit=(p3-1-1)]{};
			\node[draw, group, fill=none, fit=(p3-1-2)]{};
			\node[draw, group, fill=none, fit=(p3-1-3)]{};
			\node[draw, group, fill=none, fit=(p3-2-1)]{};
			\node[draw, group, fill=none, fit=(p3-2-2)]{};
			\node[draw, group, fill=none, fit=(p3-2-3)]{};
			\node(p3)[group, fill=none, fit=(p3-1-1)(p3-2-3)]{};
			%%%%%%%%%%%%%%%%%%%%%%%%%%%%%%%%%%%%%%%%%%%
			%
			%
			%	% clustering of Z in PLP
			%
			%	\node[draw, group, fill=none, fit=(p3-1-1)]{};
			%	\node[draw, group, fill=none, fit=(p3-1-2)]{};
			%	\node[draw, group, fill=none, fit=(p3-1-3)]{};
			%	\node[draw, group, fill=none, fit=(p3-2-1)]{};
			%	\node[draw, group, fill=none, fit=(p3-2-2)]{};
			%	\node[draw, group, fill=none, fit=(p3-2-3)]{};
			%	%
			\draw(p0)--(p1)--(p2)--(p3);
			\draw[dashed,overlay] (1)--(p0.east|-1);	
			\draw[dashed,overlay] (2)--(p0.east|-2);
			\draw[dashed,overlay] (3)--(p0.east|-3);
			\end{tikzpicture}
		}\hfill
	\end{center}
	\caption{Dilworth truncation}
\label{fig:eg-dt}
\end{figure}
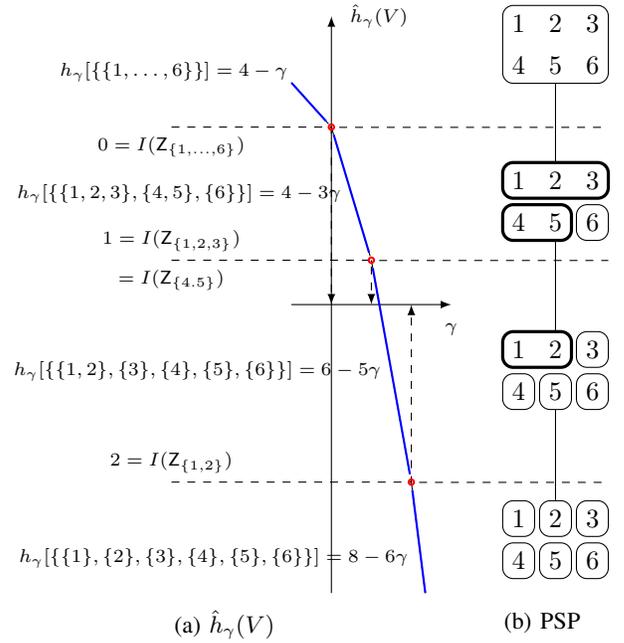

\begin{example}
%	\figref{fig:eg-dt}	
	For the motivating example, the Dillworth truncation $\hat{h}_{`g}(V)$ \eqref{eq:DT:1} is shown
	in \figref{fig:eg-dt-fn}. The Dillowrth truncation is piecewise linear with at most $|V|$
	turning points. A turning point $p_i:=(`g_i,\hat{h}_{`g_i}(V))$ occurs when \eqref{eq:DT:1} has
	more than one solution. The collection of such optimal solutions is the Dillowrth truncation
	lattice at $`g_i$ indicated in Proposition~\ref{prop:plp}. (The Dillowrth truncation lattice is
	not shown in \figref{fig:eg-dt}.) The finest optimal partition at $`g_i$ remains (while all other
	partitions seieze to be) optimal until the next turning point. In other words, the finest optimal
	partition at $`g_i$ determines the line segment that follows $`g_i$. The sequence of the finest
	optimal partitions is the PSP, whch is shown in \figref{fig:eg-psp}. 
%	each line segment
%	in the 
%	Moreove, for each 
%	lattices for all values of $`g$ are shown in
%	\figref{fig:eg-plp}. For instance, the sublattice consisting of the top five partitions in
%	\figref{fig:eg-plp} is the Dillworth trunction lattice at $`g=0$, i.e., it contains all the
%	solutions to $h_{0}[\mcP] = \hat{h}_{0}(V)$. (This can be seen from \figref{fig:eg-dtl-fn} at
%	$p_1$.) The finest 
\end{example}
%%%%%%%%%%%%%%%%%%%%%%%%%%%

\subsection{Principal sequence and minimum norm base}
\label{sec:PS}
%\input{PS}
%The computation of the PSP (and therefore the info-clustering solution) will make use of an intimately related mathematical structure called the principal sequence (PS). 

Consider any submodular function $f:2^U\to `R$~\eqref{eq:submodular} on the finite ground set $U$. For $`l\in `R$, define
\begin{align}
\kern-.5em S_{`l}(f):=\max \argmin_{B\subseteq V} f(B)-`l\abs{B}. \label{eq:S_`l}
\end{align}
where $\max$ is the inclusion-wise maximum.
Note that the minimization in \eqref{eq:S_`l} is a \emph{submodular function minimization (SFM)} since the function $B\subseteq U\mapsto f(B)-`l\abs{B}$ is also submodular. It is well-known that the minimizers form a lattice with respect to set inclusion~\cite{schrijver02}, and so the maximum in \eqref{eq:S_`l} exists and is unique.  
\begin{Proposition}[\mbox{\cite{fujishige80,fujishige-pp-revisited}}] 
	\label{prop:ps}
	$S_{`l}(f)$ for $`l\in`R$ satisfies
	 \begin{align*}
	 	S_{`l'}(f)\subseteq S_{`l''}(f) \kern1em \text {iff} \kern1em `l'\leq `l''
	 \end{align*}
	 and is referred to as the principal sequence (PS).
\end{Proposition}
%$S_{`l}(f)$ can be characterized and computed in strongly polynomial time as follows:

Without loss of generality, we assume $f$ is normalized, i.e., $f(`0)=0$, because we can redefine
$f$ as $f-f(`0)$ without  affecting the solutions to the SFM in~\eqref{eq:S_`l}, i.e. the PS is
invariant to constant shift in the submodular function. The polyhedron $\rmP(f)$ and base polyhedron
$\rmB(f)$ of the submodular function $f$ are defined as
\begin{align}
	\rmP(f)&:=\Set {x_U\in `R^U\mid x(B)\leq f(B),\forall B\subseteq U} \label{eq:rmP}\\
	\rmB(f)&:=\Set {x_U\in \rmP(f)\mid x(U)=f(U)}\label{eq:rmB}
\end{align}
where $x_U:=(x_i\mid i\in U)$ and $x(B):=\sum_{i\in B} x_i$ for convenience. ($\rmP(f)$ and
$\rmB(f)$ are non-empty as $f(`0)\geq 0$.) With $\norm {x_U}$ denoting the Euclidean norm of the
vector $x_U$, the following holds.

\begin{Proposition}[\mbox{\cite{fujishige80,fujishige11}}]
	\label{prop:MNB}
	For any normalized submodular function $f$~\eqref{eq:submodular},
	\begin{align}
		\label{eq:MNB}
		\min \Set{\norm{x_U}\mid x_U\in \rmB(f)}
	\end{align}
	has a unique solution $x_U^*$, called the minimum (Euclidean) norm base, which satisfies
	\begin{subequations}
	\begin{alignat}{2}
		\kern-1em x^*_i(f)&=\min\Set{`l\in`R\mid i\in  S_{`l}(f)}&\kern1em& \forall i\in U, \text {or equiv.,}\label{eq:MNB:x*}\kern-.5em\\
		\kern-1em S_{`l}(f)&=\Set{i\in U\mid x^*_i(f)\leq `l},&\kern1em & \forall `l\in `R,\label{eq:MNB:S_`l}
	\end{alignat}
	\end{subequations}
	where the equivalence follows directly from Proposition~\ref{prop:ps}.
\end{Proposition}
The minimum norm base may be computed using Wolfe's minimum norm point algorithm as in~\cite{fujishige11}, and so as the PS by \eqref{eq:MNB:S_`l}. Conversely, by \eqref{eq:MNB:x*}, the minimum norm base can also be computed by any SFM algorithm that solves $S_{`l}$ for any $`l$, but the minimum norm point algorithm was shown~\cite{fujishige11} empirically to perform well compared to other submodular function minimization algorithms. 

\section{Main results}
\label{sec:results}

\begin{algorithm}
	\caption{Implementation of \texttt{Fuse} in Algorithm~\ref{alg:aic}.}
	\label{algo:fuse}
	\SetKwProg{myfn}{function}{:}{end}
	\SetKwFunction{MinNormBase}{MinNormBase}
	\DontPrintSemicolon
	\SetAlgoLined
	%	\KwData{Statistics of $\RZ_V$ sufficient for calculating \FirstClusters{$B$} for all $B\subseteq V:\abs{B}>1$.}
	%	\KwResult{$\mcS$ is a list of $(I(\RZ_C),C)$ for $C\in \pzC(\RZ_V)$, which gives
	%		$\Upgamma(\RZ_V)=\Set{`g' \mid (`g',B)\in \mcS}$ and
	%		$\pzC_{`g}(\RZ_V)=\op{maximal}\Set{B\mid (`g',B)\in \mcS,`g'>`g}$.}
	%	\BlankLine
	\BlankLine
	\KwData{$\mcP$ is equal to $\mcP_{\ell}$ for some $1\leq \ell\leq N$ in Proposition~\ref{prop:clusters}.}
	\KwResult{$(`g,\mcP')$ is equal to $(`g_\ell,\mcP_{\ell-1})$.}
	%\myfn{\Fuse{$\mcP$}}{
		Enumerate $\mcP$ as $\Set{C_1,\dots,C_k}$ for some $k>1$ and disjoint $C_i$'s;\;
		\DataSty{x}$\leftarrow$ empty array of size $k$;\;
		\For{$j = 1$ \emph{\KwTo} $k$ }{
			\DataSty{x}[$j$]=\MinNormBase($ B\mapsto h`1(\bigcup_{i\in B\cup\Set{j}}C_i`2)-\sum_{i\in B\cup\Set{j}}h(C_i),\Set{j+1,\ldots,k}$);\; \label{ln:MNB:1}
		}
		%\KwRet $(\upgamma_1(\RZ_B),\pzC_{\upgamma_1}(\RZ_B))$;
		$\displaystyle`g\leftarrow -\min_{i,j:1\leq i<j\leq k}\DataSty{x}[j][i]$,\kern1em$\mcP'\leftarrow`0$;\;
		\For{$j = 1$ \emph{\KwTo} $k$ }{
			\If{$C_j\not\in \bigcup\mcP'$}{
				add $\Set{C_j}\cup\Set{C_i\mid i\kern-.25em\in\kern-.25em\Set{j\kern-.25em +\kern-.25em 1,\dots,k}, \DataSty{x}[j][i]\kern-.25em\leq\kern-.25em -`g}$ to $\mcP'$;\;
			}
			
		}
		
	%}
	\myfn{\MinNormBase($f,U$)}{
		%\tcc{ assume normalized submodular function $f:2^U\mapsto`R$ }
		\KwRet an array \DataSty{x} (indexed by $U$) that solves~\eqref{eq:MNB}.
	}
\end{algorithm}

The main result is the implementation of \Fuse in Algorithm~\ref{algo:fuse} that computes the PSP and therefore the info-clustering solution iteratively from the finer partitions to the coarser ones. This is done by computing the PS using a subroutine \MinNormBase that computes the minimum norm base that solves \eqref{eq:MNB}. An explicit implementation of this subroutine can be found in \cite{fujishige11} using Wolfe's minimum norm point algorithm. The current abstraction also allows further approximations or simplifications for special source models as in \cite{chan16cluster}. 

To explain Algorithm~\ref{algo:fuse}, we first simplify what \Fuse should compute as follows.

\begin{Theorem}
	\label{thm:`g:P:fused}
 Consider $\mcP_{\ell}$ and $`g_\ell$ defined as in Proposition~\ref{prop:clusters} for $1\leq \ell\leq N$ and write
 \begin{align*}
 	\mcP_{\ell}=\Set{C_j\mid j\in U} \kern1em \text{and}\kern1em
 	\RZ'_j:=\RZ_{C_j}
 \end{align*}
 for some index set $U$ and disjoint subset $C_j$ for $j\in U$. Then, for $1\leq \ell \leq N$ we
 have
 \begin{subequations}
 	\begin{align}
 	`g_\ell&=\max_{\mcF\subseteq \mcP_{\ell},\abs{\mcF}>1} I(\RZ_{\bigcup \mcF}) \label{eq:A:`g_l}\\
 	&=I^*(\RZ'_U)%:=\max`1\{I(\RZ'_B)\mid B\subseteq U,\abs{B}>1`2\}
 	\label{eq:A:`g_l:fused}\\
 	\nextParentEquation
 	\kern-.7em\mcP_{\ell-1}`/\mcP_{\ell}
 	&=\op{maximal}\Set*{\bigcup\mcF`1| \mcF\in\kern-.5em \argmax_{\mcF\subseteq \mcP_{\ell}:\abs{\mcF}>1} \kern-.5em I(\RZ_{\bigcup \mcF}) \kern-.3em `2.}\kern-.5em\label{eq:A:P_i-1}\\
 	&=`1\{\bigcup\nolimits_{j\in B}C_j \mid B\in \pzC^*(\RZ'_U)`2\}, \label{eq:A:P_i-1:fused}
 	\end{align}
 \end{subequations}
where $\bigcup \mcF:=\bigcup_{B\in \mcF} B$ for convenience.
More precisely, \eqref{eq:A:`g_l} and~\eqref{eq:A:P_i-1} follow more generally from the property~\eqref{eq:imunion}, while \eqref{eq:A:`g_l:fused} and \eqref{eq:A:P_i-1:fused} follow from 
the definition~\eqref{eq:mmi} of the MMI.
\end{Theorem}
\begin{Proof}
	See Appendix~\ref{sec:proof}.
\end{Proof}

It follows from~\eqref{eq:A:`g_l:fused} and~\eqref{eq:A:P_i-1:fused} that it suffices to compute $I^*$~\eqref{eq:I*} and $\pzC^*$~\eqref{eq:C*}. This can be done using a minimum norm base algorithm as follows. Define
\begin{align}
\label{eq:JT}
\begin{split}
J_{\opT}(\RZ_V)&:=I_{\Set{\Set{i}\mid i\in V}}(\RZ_V)\\
&=\frac1 {\abs{V}-1}`1[\sum_{i \in V}H(\RZ_i)-H(\RZ_V)`2],
\end{split}
\end{align}
which is called the normalized total correlation~\cite{chan15mi} as it is the same as Watanabe's total correlation~\cite{watanabe60} except for the additional normalization factor of $\frac 1{\abs {V}-1}$.

\begin{Theorem}
\label{thm:JT}
With entropy function $h$~\eqref{eq:h} for $\RZ_V$, define
\begin{align*}
	U_j &:=\Set {i\in V\mid i>j} \kern1em \text{for } j\in V,\text{ assuming $V$ is ordered.}\\
	g_j(B) &:= h(B\cup \Set {j})-\sum\nolimits_{i\in B\cup \Set {j}} h(\Set {i}) \kern1em \text {for }B\subseteq U_j,
\end{align*}
which is a normalized submodular function. Then,
\begin{subequations}
	\begin{align}
		I^*(\RZ_V)&=\max_{C\subseteq V:\abs{C}> 1} J_{T}(\RZ_C) \label{eq:maxJT}\\
		&= -\min_{j\in V} \min_{i\in U_j} x^{(j)}_i \label{eq:maxJT:x}\\ \nextParentEquation
		\pzC^*(\RZ_V)&=\op{maximal}
		\argmax_{C\subseteq V:\abs{C}> 1} J_{T}(\RZ_C)
		\label{eq:argmaxJT}\\
		&=\op{maximal} 
		`1\{`1.\Set{j^*}\cup 
		\argmin_{i\in U_{j^*}} x^{(j^*)}_i `2|`2.  \notag\\
		&\kern6.5em 
		`1.j^*\in \argmin_{j\in V} `1( \min_{i\in U_{j}} x^{(j)}_i `2)
		`2\},\label{eq:argmaxJT:x}
	\end{align}
\end{subequations}
where $x^{(j)}_{U_{j}}$ is the minimum norm base for $g_j$~\eqref{eq:MNB}.
\end{Theorem}
\begin{Proof}
	See Appendix~\ref{sec:proof}.
\end{Proof}

\eqref{eq:maxJT} and~\eqref{eq:argmaxJT} essentially eliminate the need for minimization over
partitions in calculating the MMI in~\eqref{eq:I*} and~\eqref{eq:C*}. They serve as an intermediate
step that leads to \eqref{eq:maxJT:x} and \eqref{eq:argmaxJT:x}, which relate the last critical value \eqref{eq:I*} and the second last partition
\eqref{eq:C*} to the minimum norm base. 
%\begin{Remark}
Together with Proposition~\ref{prop:MNB}, we have the complete implementation of \Fuse as shown in Algorithm~\ref{algo:fuse} using a minimum norm base algorithm.
%For efficient implementation, we can further impose $i$ in~\eqref{eq:I^*:f} to be the smallest element in $B\cup\Set{i}$ without loss of optimality, and so the minimization can be done over a smaller set of $B\subseteq \Set{j\in V\mid j> i}$. (See line~\ref{ln:MNB:1} of Algorithm~\ref{algo:fuse}.)
%\end{Remark}

%%%%%%%%%%%%%%%%%%%%%%%%%%%%%
\begin{example}
	As an illustration of Theorem~\ref{thm:JT} (and the agglomerative algorithm), consider our running example with $V =
	\Set{1,\ldots,6}$. The minimum norm base $x_{U_{j}}^{(j)}$ of $g_{j}$ is given as 
%\begin{align*}
%	x_{U_j}^{(j)} =
%	\begin{blockarray}{ccc ccl c}
%				1 &	2 & 3 & 4 & 5 & 6 \\
%					\begin{block}{(ccc ccc)c}
%					&	\circled{-2}, & -1, & -0.5, & -0.5, & 0   & j=1 \\
%					&	   & -1,  & -0.5, & -0.5, & 0   & j=2 \\
%					&	   &      & -0.5, & -0.5, & 0   & j=3 \\
%					&	   &      &      & -1,   & 0   & j=4 \\
%					&	   &      &      &      & 0   & j=5 \\
%					\end{block}
%	\end{blockarray}
%\end{align*}
\begin{align*}
	x_{U_j}^{(j)} =
	\begin{blockarray}{lcc ccl c}
		\phantom{n}1 &	2 & 3 & 4 & 5 & 6 \\
					\begin{block}{\{lcc ccc c}
					(&	\circled{-2}, & -1, & -0.5, & -0.5, & 0),   & j=1 \\
					(&	   & -1,  & -0.5, & -0.5, & 0),   & j=2 \\
					(&	   &      & -0.5, & -0.5, & 0),   & j=3 \\
					(&	   &      &      & -1,   & 0),   & j=4 \\
					(&	   &      &      &      & 0),   & j=5. \\
					\end{block}
	\end{blockarray}
\end{align*}
By \eqref{eq:maxJT:x}, we have $I^{*}(\RZ_V) = 2$ and by \eqref{eq:argmaxJT:x}, we have
$\pzC^{*}(\RZ_V) = \Set{\Set{1,2}}$.
In other words, starting with the partition into singletons in the agglomerative algorithm, the
function \Fuse returns the threshold value $2$ and the partition
$\Set{\Set{1,2},\Set{3},\Set{4},\Set{5}, \Set{6}}$.
Let $\RZ'_1 = \RZ_{\Set{1,2}}$ and $\RZ'_{i} = \RZ_{i+1}$ for $i=2,\ldots,5$.
The minimum norm base $x_{U_{j}}^{(j)}$ of $g_{j}$ (defined using the entropy function of $\RZ'_{\Set{1,\dots,5}}$) is
given as
\begin{align*}
	x_{U_j}^{(j)} =
	\begin{blockarray}{lcc cl c}
		\phantom{n}1 & 2 & 3 & 4 & 5 \\
					\begin{block}{\{lcc cc c}
							 (& \circled{-1}, & -0.5, & -0.5, & 0),   & j=1 \\
							 (&     & -0.5, & -0.5, & 0),   & j=2 \\
							 (&     &       & \circled{-1},   & 0),   & j=3 \\
							 (&     &       &       & 0),   & j=4 \\
					\end{block}
	\end{blockarray}
	\end{align*}
By \eqref{eq:maxJT:x}, we have $I^{*}(\RZ'_{\Set{1,\ldots,5}}) = 1$ and by \eqref{eq:argmaxJT:x}, we have
$\pzC^{*}(\RZ'_{\Set{1,\ldots,5}}) = \Set{\Set{1,2},\Set{3,4}}$, which in terms of
$\RZ_{V}$, results in the clusters $\Set{\Set{1,2,3},\Set{4,5}}$.
In other words, when called with the input partition 
$\Set{\Set{1,2},\Set{3},\Set{4},\Set{5}, \Set{6}}$, the function \Fuse returns the threshold value
$1$ and the partition $\Set{\Set{1,2,3},\Set{4,5}, \Set{6}}$.
Let $\RZ''_1 = \RZ_{\Set{1,2,3}}$, $\RZ''_{2} = \RZ_{\Set{4,5}}$, and $\RZ''_{3} = \RZ_{6}$.
The minimum norm base $x_{U_{j}}^{(j)}$ of $g_{j}$ (defined using the entropy function of
$\RZ''_{\Set{1,2,3}}$) is given as
\begin{align*}
	x_{U_j}^{(j)} =
	\begin{blockarray}{rc c c}
		\phantom{n}1 & 2   & 3 \\
					\begin{block}{\{l cc c}
						(& \circled{0},   & \circled{0}),   & j=1 \\
						(&     &               \circled{0}),   & j=2 \\
					\end{block}
	\end{blockarray}
	\end{align*}
By \eqref{eq:maxJT:x}, we have $I^{*}(\RZ''_{\Set{1,2,3}}) = 0$ and by \eqref{eq:argmaxJT:x}, we have
$\pzC^{*}(\RZ''_{\Set{1,2,3}}) = \Set{1,2,3}$, which in terms of
$\RZ_{V}$, results in the cluster $\Set{1,\dots,6}$.
In other words, when called with the input partition 
$\Set{\Set{1,2,3}, \Set{4,5}, \Set{6}}$, the function \Fuse returns the threshold value
$0$ and the trivial partition $\Set{1,\ldots,6}$, where at this point the agglomerative algorithm
terminates.
\end{example}
%%%%%%%%%%%%%%%%%%%%%%%%%%%%%

The complexity of the algorithm is mainly due to the computation of the minimum norm base in
line~\ref{ln:MNB:1}.
%The number of such computation is at most $\abs{V}$.
This computation is repeated at most $\abs{V}$ times.
With $\op{MNP}(l)$ being
the complexity of the minimum norm base algorithm for ground set of size $l$, then \Fuse runs in
time $O(\abs{V}\op{MNP}(\abs{V}))$. Since the agglomerative info-clustering algorithm in
Algorithm~\ref{alg:aic} invokes function \Fuse $N-1\leq \abs{V}-1$ times, it runs in time
$O(\abs{V}^2 \op{MNP}(\abs{V}))$, which is equivalent to that of~\cite[Algorithm~3]{chan16cluster},
assuming that the submodular function minimization therein is implemented by the minimum norm base
algorithm, i.e., with $\op{SFM}=\op{MNP}$. The divisive info-clustering algorithm
in~\cite[Algorithm~2]{chan16cluster} makes $N-1$ calls to a subroutine that calculates the
fundamental partition. However, computing the fundamental partition appears to take time
$O(\abs{V}^2 \op{MNP}(\abs{V}))$, which would lead to an overall complexity of $O(\abs{V}^3
\op{MNP}(\abs{V}))$ for the divisible clustering. Hence, the agglomerative info-clustering appears
more efficient.

\section{Conclusion}
\label{sec:conclusions}
To address the concern of entropy estimation and computational complexity, we have proposed an
agglomerative info-clustering approach that merges smaller clusters into larger clusters
successively. To the best of our knowledge, this is the fastest info-clustering algorithm without
any approximation. As mentioned in \cite{chan16cluster}, however, faster algorithms are possible
under special source models, such as the Markov tree model or Chow--Liu tree
approximation~\cite{chan15allerton}. For the graphical source models, the PSP can be computed more
efficiently using a parametric maxflow algorithm as in \cite{kolmogorov10}. The info-clustering
algorithm can also be used to compute the solution of some related problems such as the optimal
discussion rate tuple for successive omniscience~\cite{chan16so}.

%\input{comments3}

%\bibliographystyle{IEEEtran}
%\bibliography{IEEEabrv,ref}
%\end{document}

%\clearpage
%\newpage
\appendices

\makeatletter
\@addtoreset{equation}{section}
\renewcommand{\theequation}{\thesection.\arabic{equation}}
\renewcommand{\theparentequation}{\thesection.\arabic{parentequation}}
\@addtoreset{Theorem}{section}
\renewcommand{\theTheorem}{\thesection.\arabic{Theorem}}
\@addtoreset{Lemma}{section}
\renewcommand{\theLemma}{\thesection.\arabic{Lemma}}
\@addtoreset{Corollary}{section}
\renewcommand{\theCorollary}{\thesection.\arabic{Corollary}}
\@addtoreset{Example}{section}
\renewcommand{\theExample}{\thesection.\arabic{Example}}
\@addtoreset{Remark}{section}
\renewcommand{\theRemark}{\thesection.\arabic{Remark}}
\@addtoreset{Proposition}{section}
\renewcommand{\theProposition}{\thesection.\arabic{Proposition}}
\@addtoreset{Definition}{section}
\renewcommand{\theDefinition}{\thesection.\arabic{Definition}}
\@addtoreset{Subclaim}{Theorem}
\renewcommand{\theSubclaim}{\theLemma.\arabic{Subclaim}}
\makeatother

\section{Proofs of main results}
 \label{sec:proof}
 
\begin{Proof}[Theorem~\ref{thm:`g:P:fused}]
	First, we prove \eqref{eq:A:`g_l} and \eqref{eq:A:P_i-1} using the general property~\eqref{eq:imunion} instead of the precise definition~\eqref{eq:mmi}. 

	We first argue that, for any feasible solution $\mcF$ to the r.h.s.\ of~\eqref{eq:A:`g_l}, 
	\begin{align*}
		I(\RZ_{\bigcup\mcF})\utag{a}\leq `g_\ell.
	\end{align*}
	Suppose to the contrary that $I(\RZ_{\bigcup\mcF})>`g_{\ell}$. Then, there exists $C\supseteq\bigcup \mcF$ such that
	$C\in \pzC_{`g_\ell}(\RZ_V)$ by the definition~\eqref{eq:clusters} of clusters. However, $\pzC_{`g_\ell}(\RZ_V)\subseteq \mcP_{\ell}$ by Proposition~\ref{prop:clusters}, which contradicts the fact that $\mcF\subseteq \mcP_{\ell}$ with $\abs {\mcF}>1$. 
	
	Next, we show that \uref{a} can be achieved with equality for some feasible solution $\mcF$. Consider any $C\in \mcP_{\ell-1}`/\mcP_{\ell}$. (Such a $C$ exists since $\mcP_{\ell}$ is strictly finer than $\mcP_{\ell-1}$.) Then, we have 
	\begin{align*}
	C \utag{b}= \bigcup \mcF\text{ for some }\mcF \subseteq \mcP_{\ell}:|\mcF| > 1,
	\end{align*}
	i.e., for some feasible solution $\mcF$.
	
	By Proposition~\ref{prop:clusters}, we have 
	\begin{align*}
		C\in \pzC_{`g}(\RZ_V)\kern1em \text{ for all
		 }`g\in [`g_{\ell-1},`g_\ell),
	\end{align*}
	and so $I(\RZ_C)\geq `g_\ell$, i.e., larger than all values in the interval.
	The reverse inequality also holds by \uref{a} and \uref{b}. Hence, we have
	\begin{align*}
		I(\RZ_C)\utag{c}=`g_\ell,
	\end{align*}
	 which implies \eqref{eq:A:`g_l} as desired. 
	
	Now, we argue that the above construction gives all the optimal solutions to the r.h.s.\ of \eqref{eq:A:`g_l}, hence establishing ~\eqref{eq:A:P_i-1}. For any $C\in \mcP_{\ell-1}`/\mcP_{\ell}$, \uref{b} and \uref{c} implies that ``$\subseteq$'' holds for~\eqref{eq:A:P_i-1}, because 
	the fact that $C\in \pzC_{`g_{\ell-1}}(\RZ_V)$ (by Proposition~\ref{prop:clusters}) means that it is maximal by the definition~\eqref{eq:clusters} of  clusters. 
	
	To argue the reverse inclusion ``$\supseteq$'', consider any $\mcF$ belonging to the r.h.s.\ of~\eqref{eq:A:P_i-1}. By \eqref{eq:A:`g_l}, 
	\begin{align*}
		I(\RZ_{\bigcup \mcF})=`g_\ell>`g_{\ell-1}
	\end{align*}
	and so $\bigcup \mcF\in \pzC_{`g_{\ell-1}}$ by the definition~\eqref{eq:clusters} of clusters and the maximality of $\bigcup \mcF$. This completes the poof of~\eqref{eq:A:P_i-1}.
	
	Consider proving \eqref{eq:A:`g_l:fused}
	and~\eqref{eq:A:P_i-1:fused}.
	For any optimal solution $\mcF$ to \eqref{eq:A:`g_l},  we have
	\begin{align*}
		\pzP^*(\RZ_{\bigcup \mcF})=\mcF=\Set{C_j\mid j\in B}
		%\label{eq:a:4}
	\end{align*}
	for some $B\subseteq U$, 
	where the first equality is by Proposition~\ref{prop:fund:cluster}
	since $I(\RZ_C)>`g_\ell$ for all $C\in \mcF$ such that ${C}>1$. Hence,
	\begin{align*}
		I(\RZ_{\bigcup \mcF})=I_{\pzP^*(\RZ_{\bigcup \mcF})}(\RZ_{\bigcup \mcF})&=I_{\mcF}(\RZ_{\bigcup \mcF})\\&=I_{\Set{\Set{j}\mid j\in B}}(\RZ'_B)\geq I(\RZ'_B)
	\end{align*}
	where the inequality follows from the fact that partition $\Set{\Set{j}\mid j\in B}$ into singletons  may not be the optimal partition of $B$ for $I(\RZ'_B)$. The reverse inequality also holds because, for all $\mcP'\in \Pi'(B)$, define 
	\begin{align*}
		\mcP:=\Bigg\{\bigcup_{j\in C'}C_j\mid C'\in\mcP' \Bigg\}\in \Pi'\Big(\bigcup\mcF\Big),
	\end{align*}
	we have $I_{\mcP'}(\RZ'_B)=I_{\mcP}(\RZ_{\bigcup\mcF})\geq I(\RZ_{\bigcup \mcF})$. Here, the inequality follows from the fact that $\mcP$ may not be the optimal partition of $\RZ_{\bigcup\mcF}$.
	 This completes the proof of Theorem~\ref{thm:`g:P:fused}.
\end{Proof} 
 
\begin{Proof}[Theorem~\ref{thm:JT}]
	Consider $`g_N,\mcP_{N-1}$, and $\mcP_{N}$ as in Proposition~\ref{prop:clusters} and let $h_{`g}$
	be as in~\eqref{eq:residualH}.
	By Proposition~\ref{prop:psp}, we have for all $C\in \mcP_{N-1}`/\mcP_{N}$ that 
	%Then for all $C\in \mcP_{N-1}`/\mcP_{N}$, we have by Proposition~\ref{prop:psp} (applied to $C$) that 
	\begin{subequations}
		\begin{align*}
			h_{`g_{N}}(C)&\utag{a}=\sum_{i\in C}h_{`g_{N}}(\Set{i}) \kern1em \text{or equivalently,}\\
			`g_{N}&\utag{b}=J_{T}(\RZ_C).
		\end{align*}
	\end{subequations}
\begin{compactitem}
	\item  The equivalence between~\uref{a} and~\uref{b} follows immediately from the definition~\eqref{eq:JT} of $J_{\opT}$. More precisely, \uref{b} is equivalent to
	\begin{alignat*}{2}
	&\kern1em `g_N  =\frac1 {\abs{C}-1}`1[\sum_{i \in C}h(\Set {i})- h(C)`2] &\kern1em &\text {by \eqref{eq:h}} \\
	&\iff (\abs{C}-1) `g_N = \sum_{i \in C} h(\Set {i})- h(C) & & \because \abs {C}>1\\
	&\iff \kern1.8em h_{`g_N}(C) = \sum_{i \in C} h_{`g_N}(\Set {i}) && \text {by \eqref{eq:residualH}}
	\end{alignat*}
	which is equivalent to \uref{a} as desired.
	\item ``$\geq$'' for \uref{a}
	follows from 
	\begin{align*}
	h_{`g_{N}}\big[\Set{C}\cup\Set{\Set{i}\mid i\in V`/C}\big]
	\geq
	h_{`g_{N}}(\mcP_N)
	\end{align*}
	since $\mcP_N$ is optimal to $\hat{h}_{`g_{N}}(V)$~\eqref{eq:DT} by construction.
	\item To explain ``$\leq$'' for \uref{a}, note that $\mcP_{N-1}$ is an optimal partition to the Dilworth truncation~\eqref{eq:DT:1} for $`g\in [`g_{N-1},`g_N)$ by Proposition~\ref{prop:psp}. By continuity of $h_{`g}[\mcP_{N-1}]$~\eqref{eq:DT:2} with respect to $`g$, we have that $\mcP_{N-1}$ is also optimal for $`g=`g_N$, i.e.,
	\begin{align*}
	h_{`g_{N}}[\mcP_N] \utag{c}= h_{`g_{N}}[\mcP_{N-1}],
	\end{align*}
	by the optimality of $\mcP_N$. Hence, since
	\begin{align*}
	\sum\limits_{i\in \mcP_{N}} h_{`g_{N}}(\{i\})  = \sum\limits_{C\in \mcP_{N-1}} \sum\limits_{i\in C} h_{`g_{N}}(\{i\}),
	\end{align*}
	we have,
	\begin{align*}
	\sum_{C\in\mcP_{N-1}`/\mcP_N}`1[h_{`g_{N}}(C)-\sum_{i\in C}h_{`g_{N}}(\Set{i})`2]=0.
	\end{align*}
	Each term in the bracket is non-negative as argued before, and so must be equal to zero.
\end{compactitem}
		
	Consider any maximal solution $C'$ to the r.h.s.\ of~\eqref{eq:maxJT}. Then,
	\begin{subequations}
		\label{eq:`g_N<JT(C')}
		\begin{align*}
			`g_{N}&\utag{d}\leq J_{\op{T}}(\RZ_{C'}) \kern1em \text{or equivalently,}\\%  \label{eq:`g_N<JT(C'):1}\\
			h_{`g_{N}}(C')&\utag{e}\leq \sum_{i\in C'}h_{`g_{N}}(\Set{i}),\\% \label{eq:`g_N<JT(C'):2}
		\end{align*}
	\end{subequations}
	which follow from \uref{b} and \uref{a} respectively because the \eqref{eq:maxJT} does not require $C\in \mcP_{N-1}`/\mcP_N$. With
	\begin{align*}
		\mcP':=\Set{C'}\cup\Set{\Set{i}\mid i\in V`/C'},
	\end{align*}
	\uref{e} implies
	\begin{align*}
		h_{`g_N}[\mcP'] \leq h_{`g_N}[\mcP_N],
	\end{align*}
	which must hold with equality by the optimality of $\mcP_N$. Therefore, \uref{d} also holds with equality, and so we have \eqref{eq:maxJT} since $`g_N=I^*(\RZ_V)$. Note that, by \uref{c}, $\mcP_{N-1}$ is also an optimal partition. Furthermore, it is the largest/coarsest such partition since it is optimal for some $`g<`g_N$ and therefore larger/coarser than all optimal partitions for $`g=`g_N$ by Proposition~\ref{prop:plp}. Therefore, the set of maximal optimal solutions to the r.h.s. of \eqref{eq:maxJT}  is $\mcP_{N-1}`/\mcP_N$, which is $\pzC^*(\RZ_V)$ trivially, and so we have \eqref{eq:argmaxJT}.
	
	To prove~\eqref{eq:maxJT:x} and~\eqref{eq:argmaxJT:x}, let $`g^*=I^*(\RZ_V)$. Then, it follows from \eqref{eq:maxJT} that
	\begin{subequations}
		\label{eq:I^*:f}
		\begin{align*}
			0&=\min_{C\subseteq V:\abs{C}>1}`g^*-J_{\op{T}}(\RZ_C)\\
			&\utag{f}=\min_{C\subseteq V:\abs{C}>1}(\abs{C}-1)`g^*+h(C)-\sum_{i\in C}h(\Set{i})  \\
			&\utag{g}=\min_{j\in V}\min_{B\subseteq U_j:\abs{B}\geq1} g_j(B)+`g^*\abs{B}.
		\end{align*}
	\end{subequations}
\begin{compactitem}
	\item \uref{f} is obtained by the definition~\eqref{eq:JT} of $J_{\opT}$ and multiplying both sides of the equality by $\abs{C}-1>0$, which does neither violate the equality nor change the set of solutions;
	\item \uref{g} is obtained by changing the variable $C$ to $(j,B)$ using the bijection that sets
	\begin{align*}
		j:=\min_{i\in C} i
		\kern 1em \text {and}\kern1em B:=C`/\Set{j},
	\end{align*}
	which is possible because $\abs {C}>1$.
	We have also applied $\abs {B}=\abs {C}-1$ and the definition of $g_j$ to rewrite the expression in the minimization. %The set of solutions is preserved in the sense that $C^*$ is optimal to \eqref{eq:JT} iff 
\end{compactitem}
Consider any optimal solution $j^*$ to the R.H.S.\ of \uref{g}. Then, we have
\begin{align*}
	S_{`l}(g_{j^*}) 
	\begin{cases}
		\utag{h}\neq `0 & `l=-`g^*\\
		\utag{i}= `0 & `l<-`g^*.
	\end{cases}
\end{align*}
\begin{compactitem}
	\item \uref{h} is because, by \uref{g},
	\begin{align*}
		`0&\neq \max \argmin_{B\subseteq U_{j^*}:\abs{B}\geq 1} g_{j^*}(B)+`g^*\abs{B}\\
		&= \max \argmin_{B\subseteq U_{j^*}} g_{j^*}(B)+`g^*\abs{B}\\
		&=S_{-`g^*}(g_{j^*})
	\end{align*}
	The last equality is by the definition~\eqref{eq:S_`l} of $S_{-`g^*}$. The first equality is because allowing $B=`0$ does not change the minimum value $0$, since
	\begin{align*}
		g_{j^*}(`0)+`g^*\abs {`0}=0
	\end{align*}
	 as $g_{j^*}(`0)$. Doing so also does not affect the maximum minimizer since the new optimal solution introduced, namely $`0$, cannot be maximum trivially.
	\item To explain \uref{i}, note that \uref{g} implies for all $B\subseteq U_{j^*}:\abs {B}>1$ that
	\begin{alignat*}{2}
		g_{j^*}(B)+`g^*\abs{B} &\geq 0&\kern1em& \text {and so}\\
		g_{j^*}(B)-`l\abs{B} &\geq (-`l-`g^*) \abs {B} && \forall `l\in `R\\
		& > 0 && \forall `l<-`g^*.
	\end{alignat*}
	In other words, for $`l<-`g^*$, we have that $`0$ is the unique solution to $\min_{B\subseteq U_{j^*}} g_{j^*}(B)+`g^*\abs{B}$, which implies \uref{i} by the definition~\eqref{eq:S_`l} of $S_{`l}$.
\end{compactitem}
Now, \uref{h} and \uref{i} implies that
\begin{align*}
	-`g^*&=\sup \Set{`l\in `R\mid S_{`l}(g_{j^*})=`0}\\
	&= \min_{i\in U_{j^*}} \min_{`l\in `R:i\in S_{`l}(g_{j^*})} `l\\
	&\utag{j}= \min_{i\in U_{j^*}} x^{(j^*)}_i
\end{align*}
where the last equality \uref{j} is by \eqref{eq:MNB:x*} since $x^{(j^*)}_{U_{j^*}}$ denotes the minimum norm base for $g_{j^*}$. The equality implies \eqref{eq:maxJT:x} as desired.

To prove \eqref{eq:argmaxJT:x}, consider any set $C^*$ that belongs to the r.h.s.\ of \eqref{eq:argmaxJT}. Applying the bijection
	\begin{align*}
	j^*:=\min_{i\in C^*} i
	\kern 1em \text {and}\kern1em B^*:=C^*`/\Set{j^*},
\end{align*}
$(j^*,B^*)$ is a solution to the r.h.s.\ of \uref{g}. By the inclusion-wise maximality of $C^*$, the set $B^*$ is also a maximal (the maximum) solution, i.e.,
\begin{align*}
	B^*&\utag{k}= S_{-`g^*}(g_{j^*})\\
	&= \Set {i\in U_{j^*}\mid i\in S_{-`g^*}(g_{j^*}) }\\
	&\utag{l}= \Set {i\in U_{j^*} \mid x^{(j^*)}_i \leq -`g^*}\\
	&\utag{m}= \argmin_{i\in U_{j^*}} x^{(j^*)}_i,
\end{align*}
where \uref{k} is by \eqref{eq:S_`l}; \uref{l} is by \eqref{eq:MNB:S_`l}; and \uref{m} is by \uref{j}. Hence, $C^*$ also belongs to the r.h.s.\ of \eqref{eq:argmaxJT:x}. 

Conversely, if $(j^*,B^*)$ is an optimal solution to the r.h.s.\ of \uref{g}, it can be argued easily that $C^*:=\Set{j}^*\cup B^*$ is also a solution to the r.h.s.\ of \eqref{eq:maxJT}. The maximal such $C^*$ therefore belongs to the l.h.s.\ of \eqref{eq:argmaxJT:x} as desired. This completes the proof.
\end{Proof}

\bibliographystyle{IEEEtran}
\bibliography{IEEEabrv,ref}

\end{document}